\documentclass[aps,prb,twocolumn,showpacs,superscriptaddress]{revtex4}
\usepackage{amsfonts}
\usepackage{amsmath}
\usepackage{amssymb}
\usepackage{graphicx}

\begin{document}

\title{Phase diagram of generalized fully frustrated $XY$ model in two
dimensions}
\author{Petter Minnhagen}
\affiliation{Dept. of Physics, Ume{\aa } University, 901 87 Ume{\aa }, Sweden}
\author{Beom Jun Kim}
\affiliation{Dept. of Physics, BK21 Physics Research Division and Institute of Basic
Science, Sungkyunkwan Univ., Suwon 440-746, Korea}
\author{Sebastian Bernhardsson}
\affiliation{Dept. of Physics, Ume{\aa } University, 901 87 Ume{\aa }, Sweden}
\author{Gerardo Cristofano}
\affiliation{Dipartimento di Scienze Fisiche, Universit\'a di Napoli ``Federico II'' and
INFN, Sezione di Napoli, Via Cintia, Compl. universitario M. Sant'Angelo,
80126 Napoli, Italy}

\begin{abstract}
It is shown that the phase diagram of the two-dimensional generalized
fully-frustrated $XY$ model on a square lattice contains a crossing of the
chirality transition and the Kosterlitz-Thouless (KT) transition, as well as
a stable phase characterized by a finite helicity modulus $\Upsilon$ and an
unbroken chirality symmetry. The crossing point itself is consistent with a
critical point without any jump in $\Upsilon$, with the size ($L$) scaling $%
\Upsilon\sim L^{-0.63}$ and the critical index $\nu\approx0.77$. The KT
transition line remains continuous beyond the crossing but eventually turns
into a first-order line. The results are established using Monte-Carlo
simulations of the staggered magnetization, helicity modulus, and the
fourth-order helicity modulus.
\end{abstract}

\maketitle

\section{Introduction}
\label{sec:intro}

The phase transitions of the two-dimensional (2D)
fully-frustrated $XY$ (FF$XY$) model on a square lattice has been a subject
of controversy.~\cite{hasenbusch05r,lee91,granato91,olsson95,korshunov02} The emerging consensus
is that the model, as the temperature is lowered, first undergoes an
Ising-like transition associated with the chirality. At a slightly lower
temperature it undergoes a universal jump Kosterlitz-Thouless (KT)
transition associated with the phase angles.~\cite{olsson95,hasenbusch05r}
Since the two transitions are extremely close to each other in temperature
the question whether there is only one merged transition or actually two
separate transitions has been the cause of the controversy. An 
argument for two separate transitions with the KT transition
always at a lower temperature than the chirality transition was given by
Korshunov in Ref.~\onlinecite{korshunov02} in terms of a kink-antikink instability
of the domain walls separating domains with different chirality. The 2D
generalized fully frustrated $XY$(GFF$XY$) model has the same degrees of
freedom and the same symmetries as the FF$XY$ model. 
The argument by Korshunov~\cite{korshunov02} is quite general and
appears to hinge only on the combined $U(1)$ and $Z_2$ symmetry of
the model and the existence of Ising-like domain walls associated with the
broken $Z_2$ symmetry.~\cite{cristofano}
This strongly suggests that also the generalized model with the very same
degrees of freedom and symmetry should always have a KT transition at a 
lower temperature than the chirality transition.
{\it As shown here, this is not the case: The two transitions can merge 
in a single critical point.} The reason for this unexpected result is
that the symmetry of the model allows for a new phase 2D ``quasi'' phase-order.

The present paper is organized as follows: In Sec.~\ref{sec:GFFXY}, we introduce
the generalized fully-frustrated $XY$ model. The results of our numerical
simulations are presented in Sec.~\ref{sec:mag} for the staggered
magnetization, in Sec.~\ref{sec:helicity} for the helicity modulus,
and in Sec.~\ref{sec:Y4} for the fourth-order modulus, respectively.
Finally, Sec.~\ref{sec:summary} is devoted to the summary of the paper.

\section{The generalized FF$XY$ model} 
\label{sec:GFFXY}

The $XY$ model on a square lattice in the presence of an external magnetic
field transversal to the lattice plane is described by the action:
\begin{equation}
\label{eq:H}
H=-\frac{J}{k_BT} \sum_{\langle ij\rangle }\cos (\phi _{ij}\equiv \theta _{i}-\theta
_{j}-A_{ij}),
\end{equation}
where $\theta_i$ is the phase variable at the $i$th site, the sum is over
nearest neighbors, $J (>0)$ is the coupling constant, $T$ is the
temperature, $k_B$ is the Boltzmann constant, and 
$A_{ij} = (2e/\hbar c)\int_i^j {\bf A}\cdot d{\bf l}$ is the line
integral along the bond between adjacent sites $i$ and $j$. We consider
the case where the bond variables $A_{ij}$ are fixed, uniformly
quenched, out of equilibrium with the site variables and satisfy
the condition $\sum_p A_{ij} = 2\pi f$: here the sum is over each
set of bonds of an elementary plaquette and $f$ is the strength
of frustration. We assume that the local magnetic field in Eq.~(\ref{eq:H})
is equal to the uniform applied field; such an approximation is more valid
the smaller is the sample size $L$ compared with the transverse penetration
depth $\lambda_\perp$.~\cite{halperin} In the case of full frustration,
i.e., $f=1/2$, of interest to us here, such a model has a continuous $U(1)$
symmetry associated with the rotation of spins and an extra discrete $Z_2$
symmetry, as it has been shown by analyzing the degeneracy of the ground
state.~\cite{halsey,villain} Choosing the Landau
gauge, such that vector potential vanishes on all horizontal bonds and
on alternating vertical bounds, we get a lattice where each plaquette displays
one antiferromagnetic and three ferromagnetic bonds. Such a choice corresponds
to switching the sign of the interaction.


The generalized FF$XY$ model is obtained by changing the form of the
interaction from $-J\cos \phi $ to~\cite{domani84,jonsson94} 
\begin{equation*}
U(\phi )=\frac{2J}{p^{2}}\left[ 1-\cos ^{2p^{2}}(\phi /2)\right] .
\end{equation*}%
This does not alter any symmetry present in the original FF$XY$ model which
corresponds to $p=1$ since $2[1-\cos ^{2}(\phi /2))]=1-\cos \phi $. The
essential point is that $U(\phi )$ is periodic in $2\pi $ and that the first
term in an expansion for small $\phi $ is second order, i.e., $U(\phi
)\approx J \phi ^{2}/2$ for $\phi \ll 1$.

\begin{figure}[tbp]
\includegraphics[width=0.45\textwidth]{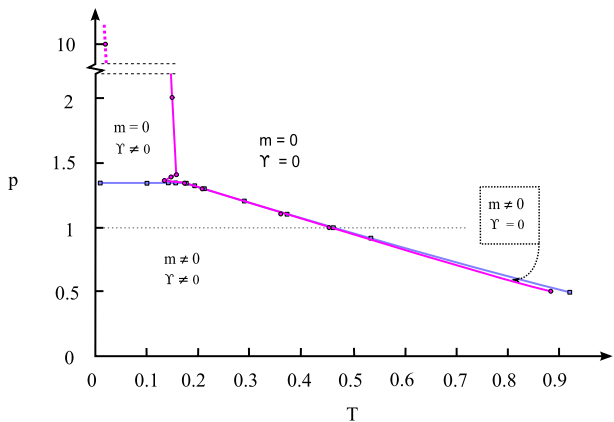}
\caption{The phase diagram of the 2D GFF$XY$ model. The phases are
characterized by the helicity modulus $\Upsilon$ and the staggered
magnetization $m$. The four phases correspond to all possible combinations
of finite and vanishing $\Upsilon$ and $m$. The points correspond to data
obtained from the simulations. The horizontal line $p=1$ corresponds to the
usual FF$XY$ model. The phase lines cross and merge in one point.}
\label{fig:phd}
\end{figure}

Figure~\ref{fig:phd} is the summary of the results present in this
paper and shows the phase diagram in the $(p,T)$-plane as
obtained from Monte-Carlo (MC) simulations.
The helicity modulus $\Upsilon$, which relates to the continuous 
angular symmetry, and the staggered magnetizations $m$,
 which relates to the discrete chirality symmetry,
are used  to detect phase boundaries.
The phase diagram
contains all four possible combinations of these two, i.e., 
$(\Upsilon,m)=(0,0),(0,\neq 0),(\neq0,0),(\neq0,\neq0)$. The dashed
horizontal line at $p=1$ corresponds to the usual FF$XY$ model, for which
the phase $(\Upsilon\neq0,m=0)$ is not realized.

\section{Staggered magnetization}
\label{sec:mag}
We first present numerical MC results of the staggered magnetization $m$, 
defined as~\cite{teitel83} 
\begin{equation*}
m= \left\langle \left| \frac{1}{L^{2}}\sum_{l=1}^{L^2} (-1)^{x_l + y_l}
s_{l} \right| \right\rangle,
\end{equation*}
where $\langle \cdots \rangle$ is the ensemble average and the vorticity for
the $l$th elementary plaquette at $(x_l, y_l)$ is computed from $s_{l}
\equiv (1/\pi)\sum_{\langle ij\rangle \in l}\phi_{ij} = \pm1$ with the sum
taken in the anti-clockwise around the given plaquette.

The ground states with the spontaneously broken chirality symmetry
correspond to the two possible checker board
patterns with alternating positive and negative vorticity. The energy per
link in these ground states is given by $U(\pi /4)$ which corresponds to all
links contributing the same energy. Since the two ground states with 
different checker board patterns are separated by an infinite energy barrier in the thermodynamic limit,
the phase with the broken chirality symmetry persists at low enough
temperatures as long as the pattern, where all links contribute the same
energy, indeed corresponds to the ground state. However, this ceases to be
true when $p$ becomes larger than $p_{c}$. In this new region the ground
state instead corresponds to a pattern consisting of plaquettes with phase
difference $0$ on three sides and $\pi $ on the remaining. The energy per
link is hence instead $U(\pi )/4$. The critical value $p_{c}$ is easily
computed to be $p_{c}\approx 1.3479$ from the condition that $U(\pi
/4)=U(\pi )/4$. Consequently, the new ground state at $p>p_{c}$ has no
broken chirality symmetry and hence corresponds to $m=0$.

\begin{figure}[tbp]
\includegraphics[width=0.45\textwidth]{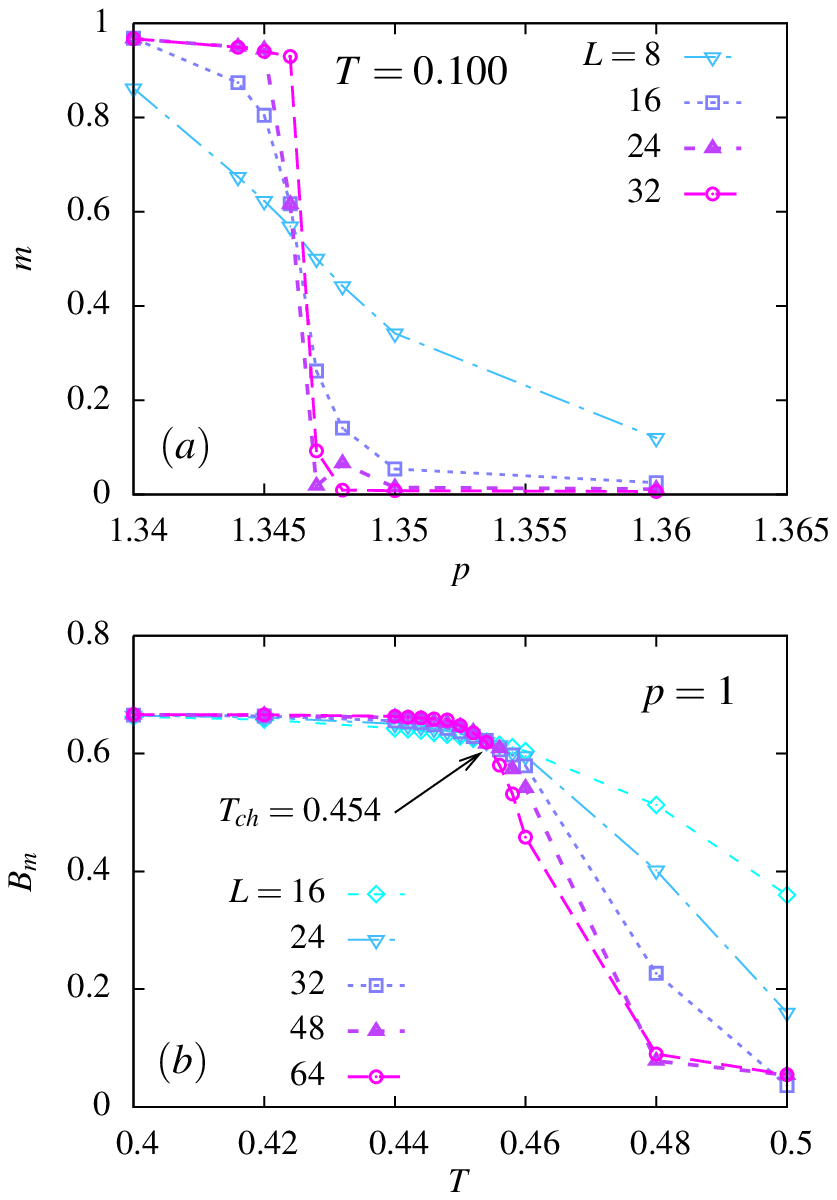}
\caption{MC determination of the staggered magnetization. (a) The transition
across the horizontal phase line $p_{c}\approx1.3479$ is confirmed by a
sharp drop to zero for larger sizes $L$. (b) Size scaling of Binder cumulant 
$B_{m}$ for $p=1$. The unique crossing of curves for different sizes yields
the transition temperature $T_{ch} \approx 0.454$. This method is used to
determine the $m$-phase line for $p < p_{c}$.}
\label{fig:mag}
\end{figure}

Figure~\ref{fig:mag}(a) illustrates the vanishing of the staggered
magnetization $m$ at $p\approx p_{c}$ for $T=0.1$ (the temperature is in
units of $J/k_{B}$ throughout the present work). This horizontal part of the 
$m$-phase boundary eventually bends down toward smaller $p$ as $T$ is
increased. This part of the $m$-phase boundary we have traced out by the
standard size scaling of Binder's cumulant $B_{m}$ for the order parameter $m
$,~\cite{binder} as displayed in Fig.~\ref{fig:mag}(b) for $p=1.$ $%
T_{ch}\approx 0.454$ is obtained, in a good agreement with the earlier value 
$0.452$ in Ref.~\onlinecite{olsson95}. The complete $m$-phase line with marked
data points are shown in Fig.~\ref{fig:phd}.

\begin{figure}[tbp]
\includegraphics[width=0.45\textwidth]{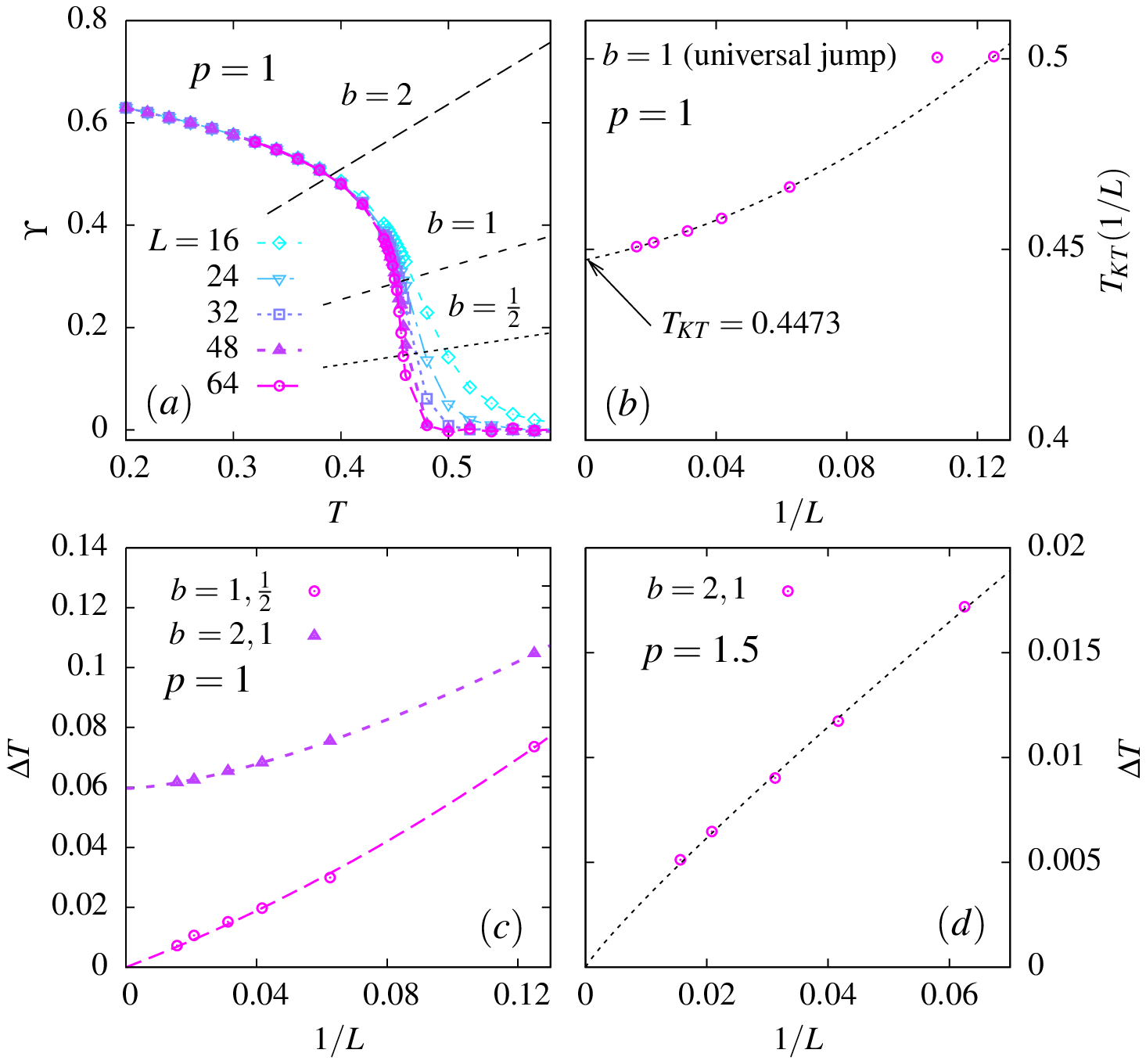}
\caption{MC determination of the KT transition line. (a) The helicity
modulus $\Upsilon$ as a function $T$ for various sizes $L$. The broken lines
correspond to the jump condition; $b=1$ is the expected universal jump. $%
T_{KT}$ is estimated by extrapolating the crossing point with the $b=1$ line
and the data to $L=\infty$, as shown in (b), which shows a second order
polynomial extrapolation of the crossing points and $T_{KT}=0.447$ is
obtained in accord with the finding $T_{KT}=0.446$ in Ref.~\protect\onlinecite%
{olsson95}. (c) The difference $\Delta T \equiv |T_{KT}(b,L) -
T_{KT}(b/2,L)| $ in crossing points $T_{KT}$ [see (a)] for $%
(b,b/2) = (1,1/2)$ and $(2,1)$. It is clearly seen that the jump has to be
less than twice the universal jump consistent with the universal jump. (d)
The same construction for $p=1.5$. We believe that a nonuniversal jump
for $p > p_c$ cannot be entirely ruled out. (a)-(c) correspond to $p=1$
while (d) is for $p=1.5$. }
\label{fig:hel}
\end{figure}

\section{Helicity modulus} 
\label{sec:helicity}
The quasi 2D phase ordering is measured by the
helicity modulus $\Upsilon $ defined as the stiffness in response to the
twist $\delta $ of the phase variables across the system: $\Upsilon \equiv
(\partial ^{2}F/\partial \delta ^{2})_{\delta =0}$ where $F$ is the free
energy. The condition for a KT transition is characterized by the universal
jump in the helicity modulus, $\Upsilon (T_{KT})/T_{KT}=2/\pi$.~\cite{nelson77,minnhagen81}. 
Thus a KT transition can be located by the
crossing point between the line $y=(2/\pi )T$ and the helicity modulus curve 
$y=\Upsilon (T)$ as illustrated for $p=1$ in Fig.~\ref{fig:hel}(a). In
practice, a precise determination requires the difficult task of
extrapolating to $L=\infty $.~\cite{hasenbusch05r} 
Here we use the following method: The values of the $T_{KT}$ at the crossing point with the
line $y=(2/\pi )T$ are determined as a function of size $L$. These values are
well approximated by a second order polynomial as shown in Fig.~\ref{fig:hel}%
(b). The extrapolation to $L=\infty $ gives $T_{KT}$ $=0.447,$ which is
very close to the value $0.446$ obtained in Ref.~\onlinecite{olsson95}. 
The close agreement shows that the method gives a good estimate of the KT transition
temperature.
The data points in Fig.~\ref{fig:phd} for $p\leq 1.32$ are obtained by this
method. One notes that the $m$-phase line and the KT-line are extremely
close for these $p$-values and only the smaller $p$-values, like $p=0.5,$
display a clear separation within our accuracy.

The determination of the KT-line rests on the assumption that the KT-jump
has the universal value $2/\pi $. A jump means that the crossing point
between the $\Upsilon (T)$ and $b(2/\pi )T$ should give the same $%
T_{KT}(L=\infty )$ for all $b\leq 1$. Figure~\ref{fig:hel}(c) shows the
difference $\Delta T(b,L)=T_{KT}(b,L)-T_{KT}(b/2,L)$ for $b=1$. Our result
is consistent with a universal jump KT transition, since $\Delta T(b=1,L)$
is consistent with a vanishing for $L=\infty $. On the other hand the jumps
size is inconsistent with a double jump since $\Delta T(b=2,L)$ approaches a
finite value. For larger values of $p>p_{c}$, like $p=1.5$, the jump is, on
the other hand, consistent with a jump larger than the universal jump as
illustrated in Fig.~\ref{fig:hel}(d): For this value $\Delta T(b=2,L)$ is
consistent with a vanishing, suggesting that the jump at the KT transition
could be larger than the universal KT value. The transition at $p=1.5$ shows
no sign of any first order character from which we conclude that it is
continuous. In this case the jump is expected to have the universal value.
Our data neither support nor rule out this expectation. Conversely, a
continuous KT transition with a nonuniversal jump can neither be ruled out.
However, when $p$ is increased further the transition does eventually become
first order as can be detected from the double well structure in the energy
histogram. For the first order transition the jump should be
nonuniversal and larger than the universal jump.~\cite{jonsson93} In the
limit of $p=\infty $ the model reduces to the infinite state Potts model,~\cite{jonsson93}
which is known to have a first order transition.

\begin{figure}[tbp]
\includegraphics[width=0.45\textwidth]{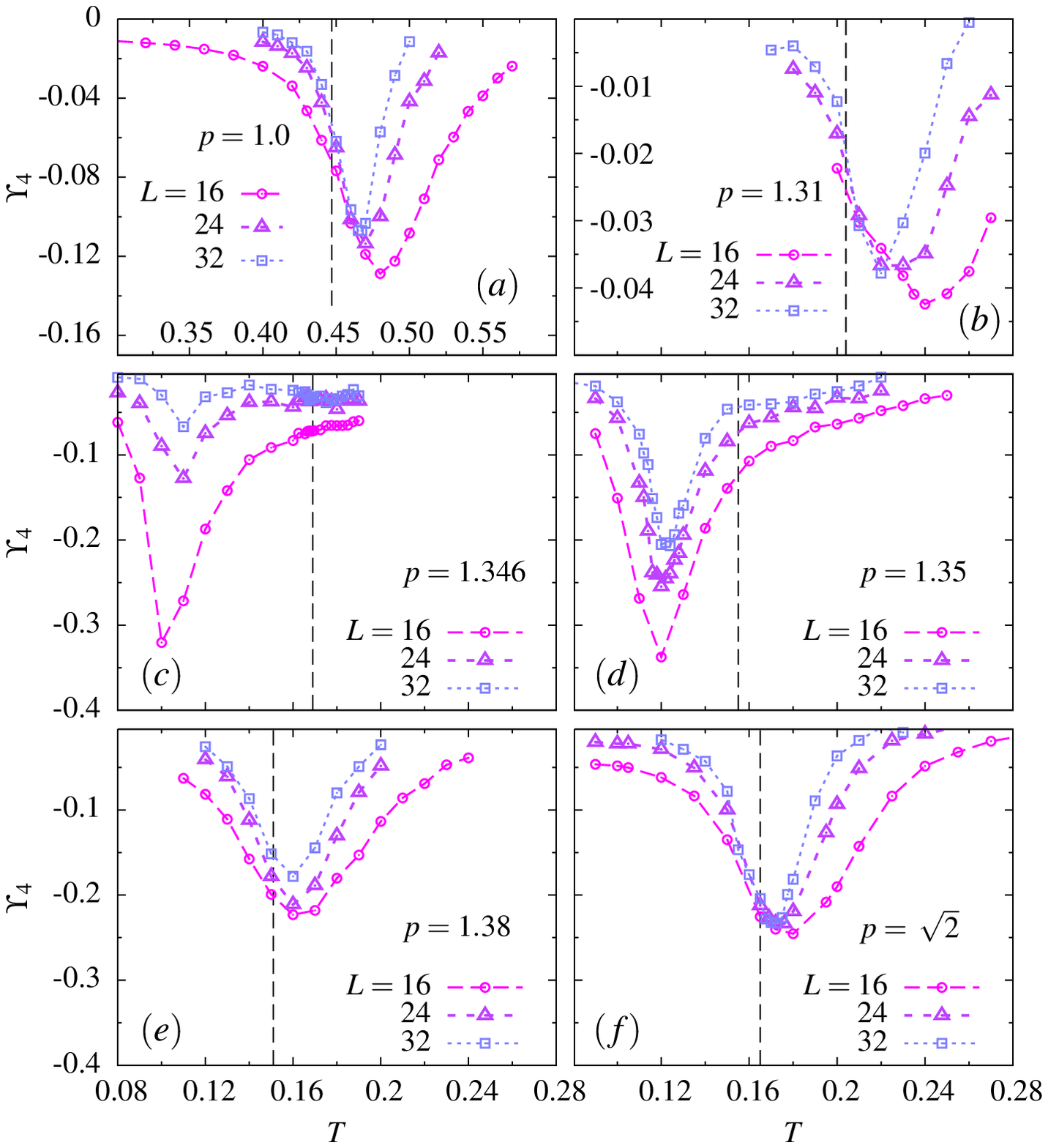}
\caption{MC simulation of the fourth-order modulus $\Upsilon_{4}.$(a) shows
the typical $\Upsilon_{4}$-characteristic of a KT-transition for $p=1$: The
well minimum moves from higher $T$ towards the transition temperature
(vertical broken line) with increasing $L$. The size of the minimum
extrapolates to a finite value. (a) and (b) show that this typical KT-feature
remains intact as $p$ is increases to the vicinity of $p_{c}$. (c) and (d)
show that this KT-feature is dramatically changed in the immediate
vicinity of $p_{c}$, while (e) and (f) show that the KT-feature is recovered
for $p$-values above $p_{c}$ }
\label{fig:y4}
\end{figure}

\section{Fourth-order modulus}
\label{sec:Y4}
According to the argument given by
Korshunov~\cite{korshunov02} the KT transition always occurs at a lower
temperature than the chirality transition.
This is consistent with the result we find for $p$-values below the horizontal
line in the phase diagram. In this part of the phase diagram the argument by
Korshunov~\cite{korshunov02} is valid and the chirality transition and the
KT transitions are separated with the KT transition always at a lower temperature.
On the other hand, when the horizontal phase line meets and crosses the KT line,
Korshunov's argument~\cite{korshunov02} is no longer valid and one expects 
a merged character of the transition. The argument by Korshunov fails because
it presumes the existence of Ising-like domain walls and such walls do not
exist above the horizontal line.
In order to monitor the change
of character of the transition we study the fourth-order helicity modulus,~\cite{minnhagen03} 
defined by the expansion of the free energy $\Delta F =
F(\delta) - F(0) =\Upsilon \delta^{2}/2! +\Upsilon_{4}\delta^{4}/4!$. The
observation that $\Upsilon_{4}$ is finite and negative precisely at the KT
transition~\cite{minnhagen03} leads to the conclusion that $\Upsilon$ 
makes an abrupt jump at $T_{KT}$ in order to fulfill the requirement of $%
\Delta F \geq 0$ at any $T$. Thus $\Upsilon_{4}$ offers a way to verify a
discontinuous jump at a KT transition.~\cite{minnhagen03}

Figure~\ref{fig:y4}(a) for $p=1$ is consistent with the typical KT-features
for $\Upsilon _{4}$: The minimum position well approaches $T_{KT}$ from
above (the vertical lines in Fig.~\ref{fig:y4} mark the expected positions
of the transition) and the depth of the well remains finite, as $L$ is
increased, confirming the existence of an abrupt jump of the helicity
modulus.~\cite{minnhagen03}
The shift of the minimum position towards lower temperatures constitutes the
characteristics of a KT transition and is well established up to $p=1.31$
[see Fig.~\ref{fig:y4}(b)]. As $p$ is increased through the crossing region
around $p_{c}\approx 1.3479$ there are dramatic changes but for larger $p$
the typical KT behavior of $\Upsilon _{4}$ reappears [see Fig.~\ref{fig:y4}%
(e)]. This is consistent with a crossing where a KT transition disappears
and reappears as $p$ is increased. The characteristics close to $%
p_{c}\approx 1.3479$ is instead consistent with $\Upsilon _{4}=0$ as $%
L\rightarrow \infty $ [see Fig.~\ref{fig:y4}(c) and (d)].

\begin{figure}[tbp]
\includegraphics[width=0.45\textwidth]{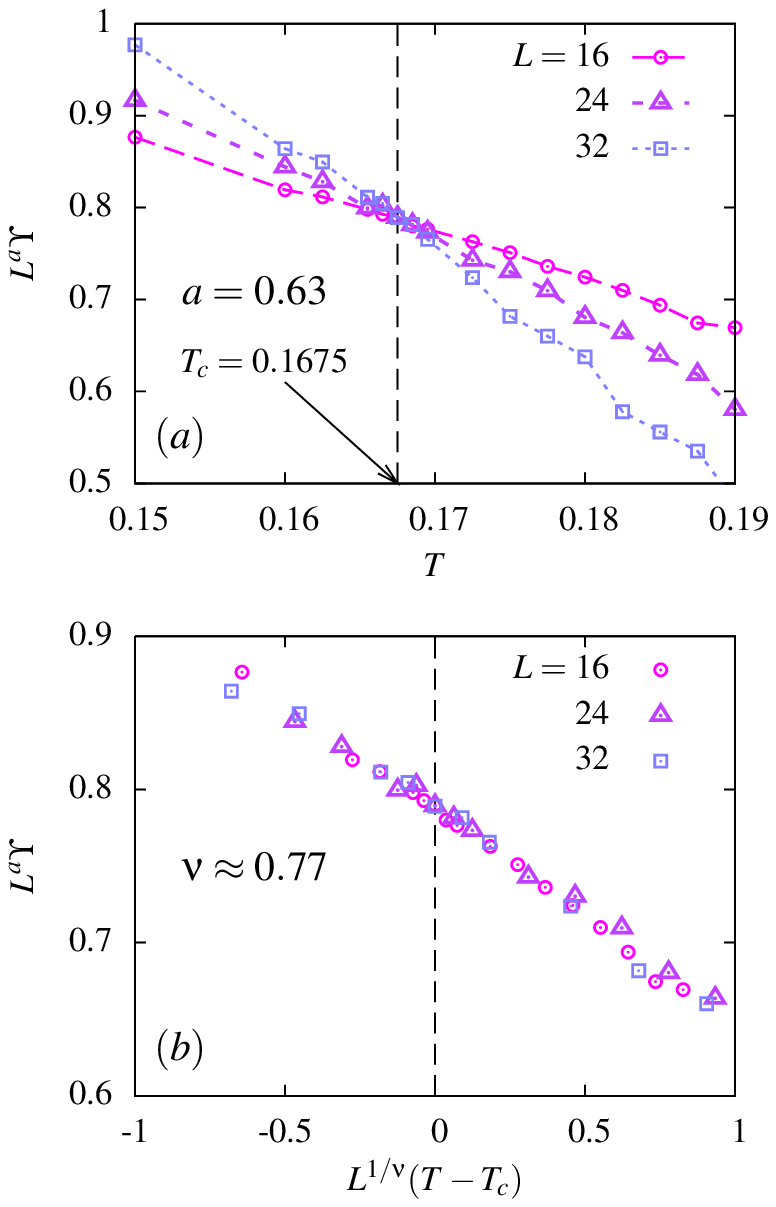}
\caption{MC-simulation of critical size scaling for the helicity modulus $%
\Upsilon$ at the crossing point of the phase lines. (a) The size scaling $%
\Upsilon\sim L^{-a}$ is obeyed to very good approximation. The collapsing
point of the data determines the critical temperature $T_{c}\approx0.1675$.
(b) With the $T_{c}$ from (a) the fuller scaling $\Upsilon=L^{-a}F[L^{1/%
\protect\nu}(T-T_c)]$ with $\protect\nu=0.77$ is born out.}
\label{fig:scale}
\end{figure}

If $\Upsilon_{4}=0$ then the helicity modulus $\Upsilon$ does not need to
have a jump at the transition, which opens up the possibility of
a continuous vanishing of $\Upsilon$ and the critical scaling
$\Upsilon\sim L^{-a}$.
Figure~\ref{fig:scale}(a) shows that
such a size scaling is indeed obtained close to $p_{c} \approx1.3479$.
Furthermore, Fig.~\ref{fig:scale}(b) shows that the standard critical
scaling form for a continuous phase transition $\Upsilon=L^{-a}F[L^{1/%
\nu}(T-T_c)]$ is also valid to very good approximation which suggests that
the correlation length $\xi$ diverges as $\xi\sim |T-T_{c}|^{-\nu}$. The
values obtained for the critical indices are $a\approx0.63$ and $%
\nu\approx0.77$.

\section{Summary} 
\label{sec:summary}
The main result of the present work is that 
in general a model with the same symmetry and degrees of freedom
as the 2D fully frustrated $XY$ model can have four stable phases.
Only three of these phases are present in the usual FF$XY$ model.
The new phase allowed by symmetry combines
unbroken
chirality with quasi 2D phase ordering. The existence of this phase also
means that the KT and chirality phase lines cross. The crossing point is a
critical point at which the helicity modulus obeys scaling and vanishes
smoothly without a universal jump.

\acknowledgments B.J.K. acknowledges the support by the Korea Research
Foundation Grant funded by the Korean Government (MOEHRD)
KRF-2005-005-J11903. P.M. and S.B. acknowledge support from the
Swedish Research Council grant 621-2002-4135.



\begin{thebibliography}{99}
\bibitem{hasenbusch05r} For a recent review, see, e.g., , M.~Hasenbusch,
A.~Peissetto, and E.~Vicari, J. Stat. Mech. 12 (2005) P12002.

\bibitem{granato91} E.~Granato, J.~M.~Kosterlitz, J.~Lee, and M.~P.
Nightingale, Phys. Rev. Lett. \textbf{66}, 1090 (1991).

\bibitem{lee91} J.~Lee, E.~Granato, and J.~M.~Kosterlitz, Phys. Rev. B 
\textbf{44}, 4819 (1991).

\bibitem{olsson95} P.~Olsson, Phys. Rev. Lett. \textbf{75}, 2758 (1995).

\bibitem{korshunov02} S.~E.~Korshunov, Phys. Rev. Lett. \textbf{88}, 167007
(2002).

\bibitem{cristofano} G. Christofano, V. Marotta, P. Minnhagen, A. Naddeo, and G. Niccoli, J. Stat. Mech. (2006) P11009.

\bibitem{halperin} B.I. Halperin, D.R. Nelson, J. Low Temp. Phys. {\bf 3}, 1165 (1979).

\bibitem{halsey} T.C. Halsey, J. Phys. C {\bf 18}, 2437 (1985); S.E. Korshunov, G.V. Uimin, J. Stat. Phys. {\bf 43}, 1
(1986).

\bibitem{villain} J. Villain, J. Phys. C {\bf 10}, 1717 (1977).

\bibitem{jonsson94} A.~Jonsson and P.~Minnhagen, Phys. Rev. Lett. \textbf{73}%
, 3576 (1994).

\bibitem{domani84} E.~Domany, M.~Schick, and R.~H.~Swendsen, Phys. Rev.
Lett. \textbf{52}, 1535 (1984).


\bibitem{teitel83} S.~Teitel and C.~Jayaprakash, Phys. Rev. B \textbf{27},
598 (1983).

\bibitem{binder} K. Binder and D.~W. Heermann, \textit{Monte Carlo
Simulation in Statistical Physics}, 2nd ed. (Springer-Verlag, Berlin, 1992).

\bibitem{minnhagen81} P.~Minnhagen, Phys. Rev. B \textbf{24}%
, 6758 (1981).

\bibitem{minnhagen03} P.~Minnhagen and B.~J.~Kim, Phys. Rev. B \textbf{67},
172509 (2003).

\bibitem{jonsson93} A.~Jonsson, P.~Minnhagen, and M.~Nyl\'en, Phys. Rev.
Lett. \textbf{70}, 1327 (1993).

\bibitem{nelson77} D.~R.~Nelson and J.~M. Kosterlitz, Phys. Rev. Lett. \textbf{%
39}, 1201 (1977).
\end{thebibliography}
\end{document}